\documentclass[journal]{IEEEtran}
\usepackage{graphicx}
\usepackage{caption,color}
\usepackage{subcaption}
\usepackage{refstyle}
\usepackage{amsthm,amssymb}
\usepackage{amsmath}
\usepackage{algorithm}
\usepackage{algorithmic}
\usepackage{epstopdf}
\usepackage{amssymb,lipsum}
\usepackage{cuted}
\ifCLASSINFOpdf
\else
\fi

\begin{document}

\title{Effective Optimization Criteria and Relay Selection Algorithms for Physical-Layer Security in Multiple-Antenna Relay Networks}

\author{Xiaotao~Lu
        and~Rodrigo~C. de~Lamare,~\IEEEmembership{Senior~Member,~IEEE}
\thanks{Xiaotao Lu is with the Communications Research Group, Department of Electronics, University of York, YO10 5DD York, U.K. e-mail: xtl503@york.ac.uk.
R. C. de Lamare is with CETUC, PUC-Rio, Brazil and with the
Department of Electronics, University of York, YO10 5DD York, U.K.
email: rcdl500@york.ac.uk.} }

\maketitle

\begin{abstract}
Physical-layer security for wireless networks has become an
effective approach and recently drawn significant attention in the
literature. In particular, the deployment and allocation of
resources such as relays to assist the transmission have gained
significant interest due to their ability to improve the secrecy
rate of wireless networks. In this work, we examine relay selection
criteria with arbitrary knowledge of the channels of the users and
the eavesdroppers. We present alternative optimization criteria
based on the signal-to-interference and the secrecy rate criteria
that can be used for resource allocation and that do not require
knowledge of the channels of the eavesdroppers and the interference.
We then develop effective relay selection algorithms that can
achieve a high secrecy rate performance without the need for the
knowledge of the channels of the eavesdroppers and the interference.
Simulation results show that the proposed criteria and algorithms
achieve excellent performance.\\
\end{abstract}


\section{Introduction}

\IEEEPARstart{I}{n} the era of modern wireless communications,
transmission security is facing great challenges due to the nature
of wireless broadcasting. To achieve transmission security in
wireless links, physical-layer security has been proposed by Shannon
\cite{Shannon} and investigated under various wireless networks
\cite{Wyner,Csiszar,Oggier}. In order to measure the security of
wireless networks, the secrecy rate (SR) has been defined as the
difference between two mutual information terms associated with the
channels of the legitimate users and the eavesdroppers,
respectively. In \cite{Oggier,Tie}, the secrecy capacity of
multi-input multi-output (MIMO) systems \cite{marzetta_first,mmimo}
is shown to be equivalent to the maximum achievable SR. More
specifically, the secrecy capacity for a particular wire-tap
Gaussian MIMO channel is determined by an achievable scheme and the
secrecy capacity region is established for the most general case
with an arbitrary number of users. Furthermore, SR optimization is
performed in the presence of a cooperative jammer.

Relay selection schemes are investigated in various scenarios such
as multiuser relay networks, cooperative relay systems and cognitive
relay networks. Relay selection methods are often considered with
the impact of co-channel interference
\cite{mmimo,Keke1,Keke2,Keke3,Xiaotao1,Xiaotao2,Costa,delamare_ieeproc},
\cite{Lee,Jingchao,aggarwal,wence,shepard,gao,ngo,hang,vardhan,li,choi,peng_twc,spa,jio_mimo,schenk,buzzi,chevalier1,chevalier2,hakkarainen,wang},
\cite{TDS_clarke,TDS_2,peng_twc,switch_int,switch_mc,smce,TongW,jpais_iet,TARMO,zlatanov,badstc,baplnc,wencewl},
\cite{Tomlinson,dopeg_cl,peg_bf_iswcs,gqcpeg,peg_bf_cl,mcpeg,Harashima,mbthpc,zuthp,rmbthp,sr_mmse,Hochwald,BDVP},
\cite{delamare_mber,rontogiannis,delamare_itic,stspadf,choi,stbcccm,FL11,P.Li,jingjing,did,bfidd}
,\cite{mbdf,jicrs}. The performance in terms of secrecy rate can be
significantly affected by the relay selection criterion adopted.
Existing relay selection algorithms depend on the knowledge of the
channels between the source to the relays and the relays to the
users \cite{Oohama}. Taking the channels from the source to the
eavesdroppers into account, a relay selection approach denoted
max-ratio criterion has been proposed in \cite{Gaojie} based on
knowledge of the channels to both legitimate users and
eavesdroppers. In prior work, the assumption of knowledge of the
channels to the eavesdroppers has been adopted even though it is
impractical. Studies have considered the max-ratio relay selection
policy, which employs the signal-to-interference-plus-noise ratio
(SINR) as the relay selection criterion and requires the knowledge
of the interference between users and the channels to the
eavesdroppers.

In this work, we examine the SR performance of multiple-relay
selection algorithms based on the SINR and the SR criteria, which
require the knowledge of the interference and the channels to the
eavesdroppers in multiuser multiple-antenna relay networks. We
present alternative optimization criteria based on the SINR and the
SR criteria that can be used for resource allocation. We then
develop novel effective relay selection algorithms based on the SINR
and SR criteria that do not require knowledge of the channels of the
eavesdroppers and interference by exploiting linear algebra
properties and a simplification of the expressions. We also analyze
the computational cost required by the proposed criteria for relay
selection and their implication on the improvement of the secrecy
rate of multiuser multiple-antenna relay networks. The SR
performance of the proposed relaying criteria and relay algorithms
is shown via simulations to approach that of techniques with full
knowledge of the interference and the channels to the eavesdroppers.

In summary, the main contributions of this work are:
\begin{itemize}
\item{Optimization criteria based on the SINR and the
SR criteria that can be used for resource allocation without the
need for the knowledge of the interference and the channels to the
eavesdroppers.}
\item{Effective relay selection algorithms based on the SINR and SR
criteria, which do not need the knowledge of the interference and
the channels to the eavesdroppers.}
\item{Analysis of the computational cost and the impact of the
proposed criteria and algorithms on the secrecy rate.}
\item{A simulation study of the proposed criteria and relay selection
algorithms in several scenarios of interest in multiuser
multiple-antenna relay networks.}
\end{itemize}

This paper is organized as follows. In Section II, the system model
is introduced. The relay selection criteria are introduced in
Section III, whereas the proposed algorithms are developed in
Section IV. The simulations are presented and discussed in Section
V. The conclusions are given in Section VI.

Notation: Bold uppercase letters ${\boldsymbol A}\in
{\mathbb{C}}^{M\times N}$ denote matrices with size ${M\times N}$
and bold lowercase letters ${\boldsymbol a}\in {\mathbb{C}}^{M\times
1}$ denote column vectors with length $M$. Conjugate, transpose, and
conjugate transpose are represented by $(\cdot)^\ast$, $(\cdot)^T$
and $(\cdot)^H$, respectively; $\boldsymbol I_{M}$ is the identity
matrix of size $M\times M$; $\rm diag \{\boldsymbol a\}$ denotes a
diagonal matrix with the elements of the vector $\boldsymbol a$
along its diagonal; $\mathcal{CN}(0,\sigma_{n}^{2})$ represents
complex Gaussian random variables with independent and identically
distributed ($i.i.d$) entries with zero mean and variance equal to
$\sigma_{n}^{2}$. $\log (\cdot)$ denotes the base-2 logarithm of the
argument.

\section{System Model}

\begin{figure}[ht]
\centering
\includegraphics[width=1.0\linewidth]{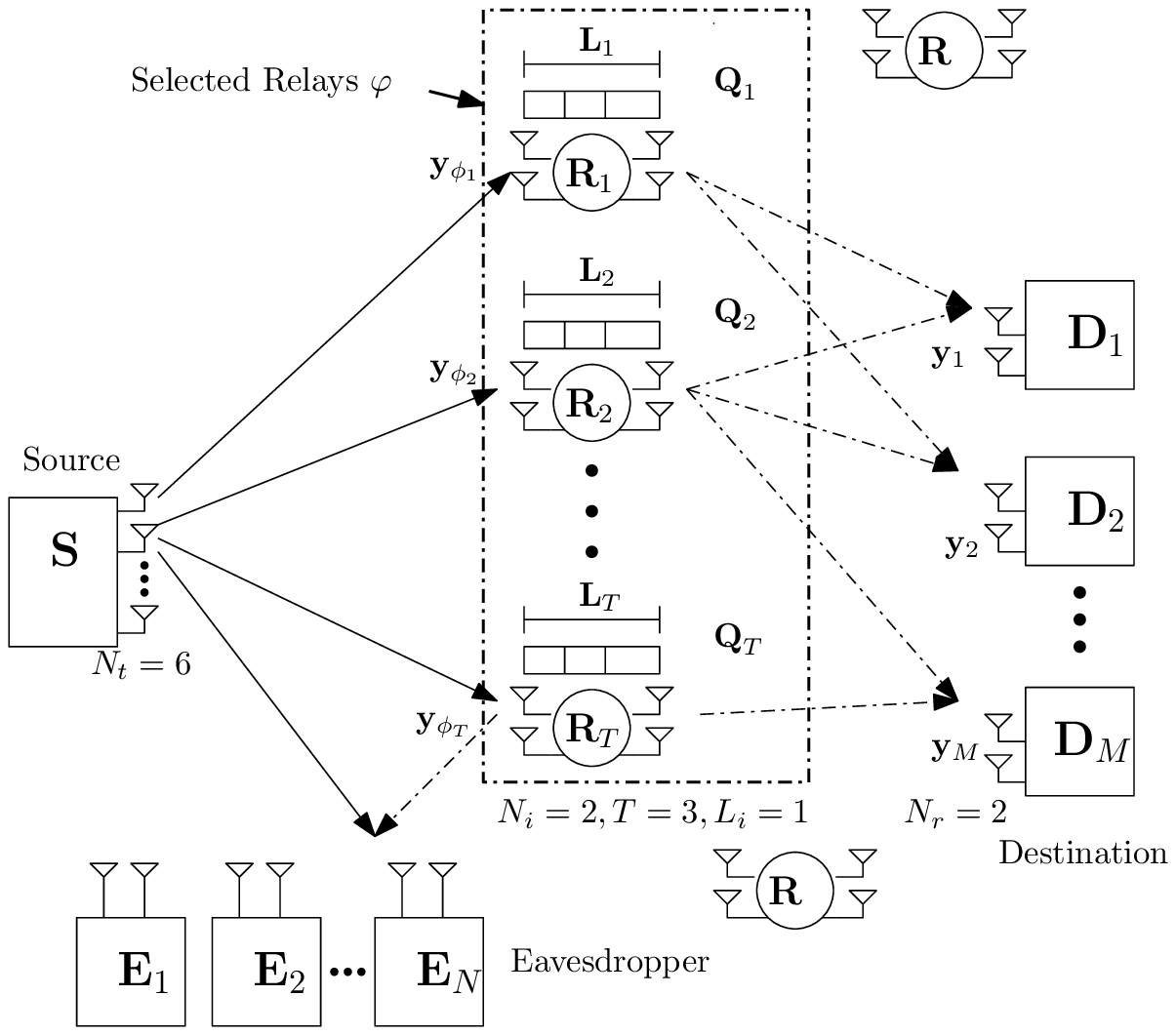}
\caption{\small Multiuser multiple-antenna network with the source
device equipped with $N_t$ antenna elements, $T$ relay devices
equipped with $N_i$ antennas each, the destination user devices with
$N_r$ antenna elements each and eavesdroppers.} \label{fig_sim}
\end{figure}

A description of the downlink multiuser multiple-antenna relay
network considered in this work is illustrated in Fig.
\ref{fig_sim}, where the system employs two time slots to transmit
the data from the source node to the users. Each relay and each user
are equipped with $N_{i}$ and $N_{r}$ antennas, respectively. We
consider a source node, which transmits $\boldsymbol s
=[{\boldsymbol s}_{1}^{T}\quad {\boldsymbol s}_{2}^{T} \quad \cdots
\quad {\boldsymbol s}_{M}^{T} ]^{T}\in {\mathbb{C}}^{MN_{r}\times
1}$ to relays. To transmit the signal simultaneously to $M$ users,
the total number of transmit antennas should be limited according to
$N_{t}^{\rm total}\geqslant MN_{r}$. For convenience, we assume that
the number of the active antennas used for transmitting user signals
is $N_{t}$ and $N_{t}=MN_{r}$. At the same time, in order to receive
the signals with a set of $T$ relays, $T \times N_{i}= N_{t}$
antennas at relays are employed. In the second time slot, the relays
will forward the signals to $M$ users. During the transmission from
the source to the users, there are $N$ eavesdroppers, which attempt
to decode the signals. Each eavesdropper is equipped with $N_{e}$
antennas.

In this system, we assume that the eavesdroppers do not jam the
transmission and the data transmitted to each user, relay, jammer
and eavesdropper experience a flat-fading multiple-antenna channel.
The source node has knowledge of the channels from the source to the
relays as well as from the relays to the users. The quantities
${\boldsymbol H}_{\phi_i}\in {\mathbb{C}}^{N_{i}\times N_{t}}$ and
${\boldsymbol H}_{e_k}\in {\mathbb{C}}^{N_{e}\times N_{t}}$ denote
the channel matrices of the $i$th relay and the $k$th eavesdropper,
respectively. If we assume $\boldsymbol \Psi$ contains sets of
$T$-combinations of the total relay set $\boldsymbol \Omega$ then
the task of relay selection is to choose the set of relays that
satisfies a chosen criterion. Given the set of selected relays
expressed as $\boldsymbol \varphi=[\phi_{1}\quad \phi_{2}\quad
\cdots \quad \phi_{T}]\in \boldsymbol \Psi$, the channel from the
transmitter to the relays and the eavesdroppers can be obtained as
$\boldsymbol H_{i}=[{\boldsymbol H}_{\phi_{1}}^{T}\quad {\boldsymbol
H}_{\phi_{2}}^{T} \quad \cdots \quad {\boldsymbol H}_{\phi_{T}}^{T}
]^{T}\in {\mathbb{C}}^{TN_{i}\times N_{t}}$ and $\boldsymbol
H_{e}=[{\boldsymbol H}_{e_{1}}^{T}\quad {\boldsymbol H}_{e_{2}}^{T}
\quad \cdots \quad {\boldsymbol H}_{e_{K}}^{T} ]^{T}\in
{\mathbb{C}}^{KN_{e}\times N_{t}}$. The matrix ${\boldsymbol
H}_{\phi_{i}r}\in {\mathbb{C}}^{N_{r}\times N_{i}}$ represents the
channel between the $i$th relay and the $r$th user. The channels
from the selected relays to the $r$th user can be described by
\begin{equation}
\boldsymbol{H}_{r}=[{\boldsymbol H}_{\phi_{1}r}\quad {\boldsymbol H}_{\phi_{2}r}
\quad \cdots \quad {\boldsymbol H}_{\phi_{T}r} ],
\end{equation}
where $\boldsymbol H_{r} \in {\mathbb{C}}^{N_{r}\times TN_{i}} $ and
$\phi_i$ represents the $i$th selected relay with a chosen relay
selection criterion. In the following section we will further
discuss various relay selection criteria, where $T$ is the total
number of selected relays
\cite{scharf,bar-ness,pados99,reed98,hua,goldstein,santos,qian,delamarespl07,xutsa,delamaretsp,kwak,xu&liu,delamareccm,wcccm,delamareelb,jidf,delamarecl,delamaresp,delamaretvt,jioel,delamarespl07,delamare_ccmmswf,jidf_echo,delamaretvt10,delamaretvt2011ST,delamare10,fa10,lei09,ccmavf,lei10,jio_ccm,ccmavf,stap_jio,zhaocheng,zhaocheng2,arh_eusipco,arh_taes,dfjio,rdrab,dcg_conf,dcg,dce,drr_conf,dta_conf1,dta_conf2,dta_ls,song,wljio,barc,jiomber,saalt}.

In Phase I, the signal is transmitted from the source to the relays.
If a precoder matrix $\boldsymbol U=[{\boldsymbol U}_{1}\quad
{\boldsymbol U}_{2} \quad \cdots \quad {\boldsymbol U}_{M}]\in
{\mathbb{C}}^{N_{t}\times MN_{r}}$ is applied, when the relay is
selected, the received signal in all relays can be expressed as
\begin{equation}
\boldsymbol y_{i}=\boldsymbol H_{i}\boldsymbol U \boldsymbol
s+\boldsymbol n_{i} \in {\mathbb{C}}^{TN_{i}\times 1},
\label{eqn:yi1}
\end{equation}
where $\boldsymbol n_{i}=[\boldsymbol n_{\phi_{1}} \quad \boldsymbol
n_{\phi_{2}}\quad \cdots \quad \boldsymbol n_{\phi_{T}}]\in
{\mathbb{C}}^{TN_{i}\times 1}$ is the noise vector that is assumed
to be  Gaussian. If the interference for relay $i$ is described by
$\boldsymbol H_{\phi_{i}}\sum_{j\neq i,j=1}^{T}{\boldsymbol
U_{j}\boldsymbol s_{j}}$, then the received signal at relay $i$ is
given by
\begin{equation}
\boldsymbol y_{\phi_i}=\boldsymbol H_{\phi_{i}}\boldsymbol U_{i}\boldsymbol
s_{i}+\boldsymbol H_{\phi_{i}}\sum_{j\neq
i,j=1}^{T}{\boldsymbol U_{j}\boldsymbol s_{j}}+\boldsymbol n_{\phi_i}. \label{eqn:yi2}
\end{equation}
In Phase II, the signal at the relay nodes is given by $\boldsymbol
y_i=[{\boldsymbol y}_{\phi_{1}}^{T}\quad {\boldsymbol
y}_{\phi_{2}}^{T} \quad \cdots \quad {\boldsymbol y}_{\phi_{T}}^{T}
]^{T}\in {\mathbb{C}}^{TN_{i}\times 1}$.

The received signal at user $r$ from the selected set of relays is
described by
\begin{equation}
\boldsymbol y_{r}=\boldsymbol H_{r}\boldsymbol y_{i}+\boldsymbol
n_{r}~ \in {\mathbb{C}}^{N_r \times 1}, \label{eqn:rsyr}
\end{equation}
where $\boldsymbol H_{r} \in {\mathbb{C}}^{N_r \times TN_{i}}$
represents the channel from the selected set of relays to user $r$
and $\boldsymbol n_{r} \in {\mathbb{C}}^{N_r \times 1}$ is the noise
vector at user $r$.

\section{Problem Statement and Relay Selection Criteria }

In this section, we state the relay selection problem and present a
generic multiple-relay selection algorithm, which relies on
exhaustive searches and can be employed with arbitrary criteria. We
then review several relay selection criteria that are available in
the literature for the system under consideration.

\subsection{Relay Selection Problem}

In the presence of multiple relay nodes, relay selection is
performed before transmission to the relays. In a half-duplex
system, we use $\eta_{1}$ to represent a metric obtained with the
information from the source to the relays and $\eta_{2}$ as another
metric calculated with information from the relays to the users.
Therefore, the relay selection criterion depends on $\eta_1$ and
$\eta_2$ and can be expressed as
\begin{equation}
\boldsymbol \varphi^{\rm select}=\rm{arg} \max_{ \boldsymbol
\varphi} \left( \eta_{1},\eta_{2}\right) \label{eqn:rs1}
\end{equation}


\begin{algorithm}
\caption{Relay selection algorithm with channel information}
\label{alg:rsch}
\begin{algorithmic}[1]
\STATE $\boldsymbol{\Omega}^{0}=\boldsymbol{\Omega}$ \STATE
$\phi^{\rm total}={\rm length}(\boldsymbol{\Omega})$ \STATE $Q={\rm
zeros}(1:\phi^{\rm total})$ \FOR {$k=1:T$} \FOR {$i=1:\phi^{\rm
total}$} \IF{$Q(i)=0$} \STATE $\theta_{\phi_i}= {\rm
trace}({\boldsymbol H}_{\phi_{i}}{\boldsymbol H}_{\phi_{i}}^{H})$
\label{op1} \STATE $Q(i)=Q(i)+1$ \ELSE \STATE$\theta_{\phi_i}=0$
\ENDIF \ENDFOR \STATE ${\boldsymbol \varphi}_{k}^{\rm select}={\rm
arg} \max_{\phi_i \in \boldsymbol{\Omega}^{0}} \{\theta_{\phi_i}\}$
\STATE ${\eta_k}=\theta_{\varphi_k^{\rm select}}$ \STATE
$\boldsymbol{\Omega}^{0}=[1 \quad 2 \quad \cdots \quad {\phi_i}-1
\quad {\phi_i}+1 \cdots  \phi^{\rm total}]$ \ENDFOR \STATE
$\eta_{1}=\sum_{k=1}^{T}{\eta_{k}}$ \STATE
$\boldsymbol{\Omega}^{1}=\boldsymbol{\Omega}$ \FOR {$r=1:M$} \FOR
{$i=1:\phi^{\rm total}$} \IF{$Q(i)\neq 0$} \STATE $\theta_{\phi_i}=
{\rm
trace}({\boldsymbol{H}_{\phi_{i}r}}{\boldsymbol{H}_{\phi_{i}r}}^{H})$
\label{op2} \STATE $Q(i)=Q(i)-1$ \ELSE \STATE$\theta_{\phi_i}=0$
\ENDIF \ENDFOR \STATE ${\boldsymbol \varphi}_{r}^{\rm select}={\rm
arg} \max_{\phi_i \in \boldsymbol{\Omega}^{1}} \{\theta_{\phi_i}\}$
\STATE ${\eta_r}=\theta_{\varphi_r^{\rm select}}$ \STATE
$\boldsymbol{\Omega}^{1}=[1 \quad 2 \quad \cdots \quad {\phi_i}-1
\quad {\phi_i}+1 \cdots  \phi^{\rm total}]$ \ENDFOR \STATE
$\eta_{2}=\sum_{r=1}^{M}{\eta_{r}}$ \STATE $\boldsymbol \varphi^{\rm
select}=\rm{arg} \max_{\boldsymbol \varphi} \left(
\eta_{1},\eta_{2}\right)$ \label{op3}
\end{algorithmic}
\end{algorithm}
A multiple-relay selection algorithm for this scenario is shown in
Algorithm~\ref{alg:rsch}. More specifically, step~\ref{op3} takes
the channel gain as the selection criterion and it can be replaced
with different selection criteria. Depending on the choice of relay
selection a designer must alter steps~\ref{op1}, ~\ref{op2} and
~\ref{op3}.

\subsection{Max-ratio criterion}
Conventional relay selection is based on the full channel
information between the source to the relays and the relays to the
users. A max-link relay selection is developed based on the max-min relay selection for decode-and-forward
(DF) relay systems \cite{Krikidis}. With the consideration of the eavesdropper, a
max-ratio selection \cite{Gaojie} in a single-antenna scenario is given by
\begin{equation}
 \phi^{\rm{max-ratio}}=\rm{arg} \max_{\phi_i \in \boldsymbol{\varphi}}\left(
\eta_{1}^{\rm{max-ratio}},\eta_{2}^{\rm{max-ratio}}\right),
\label{eqn:rsmaxr}
\end{equation}
where
\begin{equation}
\eta_{1}^{\rm{max-ratio}}=\frac{\max_{\phi_i \in \boldsymbol{\varphi} :Q{(\phi_{i})}\neq
L}\|h_{S,\phi_{i}}\|^{2}}{\|h_{se}\|^{2}}
\label{eqn:rseta1}
\end{equation}
and
\begin{equation}
\eta_{2}^{\rm{max-ratio}}=\max_{\phi_i \in \boldsymbol{\varphi} :Q{(\phi_{i})}\neq
0}\frac{\|h_{\phi_{i},D}\|^{2}}{\|h_{\phi_{i}e}\|^{2}}. \label{eqn:rseta2}
\end{equation}
Furthermore, in the scenario with only statistical distribution of the CSI to the eavesdroppers, the parameter of channel coefficient in (\ref{eqn:rseta1}) and (\ref{eqn:rseta2}) are replaced by statistical values.

\subsection{SINR criterion}
Based on the max-ratio criterion, when we consider a multiuser MIMO
system, the interference is taken into account in the relay
selection criterion. According to (\ref{eqn:yi2}), the SINR criterion can be expressed similarly as
\begin{equation}
\boldsymbol \varphi^{\rm{SINR}}=\rm{arg} \max_{\boldsymbol \varphi}\left(
\eta_{\rm{SINR}_{1}},\eta_{\rm{SINR}_{2}}\right), \label{eqn:scr}
\end{equation}
where $\eta_{\rm{SINR}_{1}}$ is the average value of SINR over relay set $\boldsymbol \varphi$. The SINR of the selected relay node $\phi_i$ can be obtained by
\begin{equation}
\begin{split}
\eta_{\varphi_{\phi_i}}^{\rm{SINR}}&={\max_{i} \rm \overline{SINR_{i}}}\\
&=\max_{i}\frac{1}{N_{i}}\sum_{n=1}^{N_{i}} \rm SINR_{n}\\
&=\max_{i}\frac{1}{N_{i}}\sum_{n=1}^{N_{i}} \rm \frac{\boldsymbol h_{\phi_{in}}^{H}\boldsymbol{R}_{d}\boldsymbol h_{\phi_{in}}}{\boldsymbol h_{\phi_{in}}^{H}\boldsymbol{R}_{In}\boldsymbol h_{\phi_{in}}}.
\label{eqn:sceta1}
\end{split}
\end{equation}
in (\ref{eqn:sceta1}), $\boldsymbol h_{\phi_{in}} \in {\mathbb{C}}^{N_{t}\times 1}$ represents the $n$th stream for the $i$th relay. $\boldsymbol R_{d}=\boldsymbol U_{i}\boldsymbol s_{i}\boldsymbol s_{i}^{H}\boldsymbol U_{i}^{H}\in {\mathbb{C}}^{N_{t}\times N_{t}}$ is
the correlation matrix of the received signal for node $i$ and $\boldsymbol R_{\rm
In}=\sum_{j\neq i,j=1}^{m}\boldsymbol U_{j}\boldsymbol s_{j}\boldsymbol s_{j}^{H}\boldsymbol U_{j}^{H}+\boldsymbol n_{j}\boldsymbol n_{j}^{H}\in {\mathbb{C}}^{N_{t}\times N_{t}}$ represents the sum of the correlation matrix of the interference and correlation matrix of the noise.
Similarly, $\eta_{\rm{SINR}_{2}}$ is the average value of SINR with relay set $\boldsymbol \varphi$ to users. The SINR of user $r$ can then be calculated by
\begin{equation}
\eta_{\boldsymbol \varphi, r}^{\rm{SINR}}=\max_{r}\frac{1}{N_{r}}\sum_{n=1}^{N_{r}} \rm \frac{\boldsymbol h_{\phi_{in}r}^{H}\boldsymbol{R}_{r}\boldsymbol h_{\phi_{in}r}}{\boldsymbol h_{\phi_{in}r}^{H}\boldsymbol{R}_{In}\boldsymbol h_{\phi_{inr}}}, \label{eqn:sceta2}
\end{equation}
where $\boldsymbol R_{r}=\boldsymbol y_{i}\boldsymbol y_{i}^{H} \in {\mathbb{C}}^{TN_{i}\times TN_{i}}$ and $\boldsymbol h_{\phi_{in}r} \in {\mathbb{C}}^{TN_{i}\times 1}$ is the $n$th stream for user $r$.

\subsection{Secrecy rate criterion}
In a multi-user MIMO system,
the secrecy-rate criterion for a multi-user MIMO system considering interferences \cite{spencercapacity} is given by
\begin{equation}
\boldsymbol{\varphi}=\rm{arg} \max_{\phi_i \in \boldsymbol \varphi,\boldsymbol{\varphi} \in \boldsymbol{\Psi}} \left\{
\log[\det({\boldsymbol{I}+\boldsymbol{\Gamma}_{r,i}})]-\log[\det({\boldsymbol{I}+\boldsymbol{\Gamma}_{e,i}})]\right\},
\label{eqn:sinrct}
\end{equation}
where $\boldsymbol{\Gamma}_{r,i}$ is given as
\begin{equation}
\boldsymbol{\Gamma}_{r,i} = (\boldsymbol H_{i}\boldsymbol R_{\rm
In}\boldsymbol H_{i}^{H})^{-1}(\boldsymbol H_{i}\boldsymbol
R_{d}\boldsymbol H_{i}^{H}), \label{eqn:rsRR2}
\end{equation}
and
\begin{equation}
\boldsymbol{\Gamma}_{e,i} = (\boldsymbol H_{e}\boldsymbol R_{\rm
In}\boldsymbol H_{e}^{H})^{-1}(\boldsymbol H_{e}\boldsymbol
R_{r}\boldsymbol H_{e}^{H}), \label{eqn:rsRE2}
\end{equation}
In (\ref{eqn:sinrct}), the criterion is based on the secrecy rate
related to destination $i$ and in a relay system the destination
can be relays as well as users.

\section{Proposed relay selection algorithms}
In the aforementioned relay selection criteria, the channels of the
source to eavesdroppers as well as the interference are assumed to be
available at the transmitter. However, this assumption is hard to
achieve in wireless transmissions \cite{Xiang}. To obviate this need, we propose effective relay selection algorithms with partial channel information.


\subsection{Simplified SINR-Based (S-SINR) Relay Selection}


The SINR selection criterion with full channel information is
expressed in (\ref{eqn:scr}). The interference signal is used to
obtain the expressions in (\ref{eqn:sceta1}) and (\ref{eqn:sceta2}).
In the following, we choose the SINR criterion and develop a
simplified SINR-based (S-SINR) relay selection algorithm with
consideration of only the channels of the users, which can be
readily obtained via feedback channels. If at the transmitter, a
linear precoder $\boldsymbol U$ is applied, the received signal can
be expressed as
\begin{equation}
\boldsymbol h_{\phi_{\rm in}}^{H}\boldsymbol R_{d}\boldsymbol h_{\phi_{\rm in}} =
\boldsymbol h_{\phi_{\rm in}}^{H}\boldsymbol U_{i} \boldsymbol s_{i}\boldsymbol
s_{i}^{H}\boldsymbol U_{i}^{H}\boldsymbol h_{\phi_{\rm in}}. \label{eqn:prd1}
\end{equation}
With linear zero-forcing precoding, we have $\boldsymbol h_{\phi_{\rm in}}^{H}\boldsymbol U_{i}=[1\quad0\quad \cdots \quad 0]$ and (\ref{eqn:prd1}) can be
written as
\begin{equation}
\boldsymbol h_{\phi_{\rm in}}^{H}\boldsymbol R_{d}\boldsymbol h_{\phi_{\rm in}}= \sigma_{d}^{2}, \label{eqn:prd2}
\end{equation}
where $\boldsymbol R_{d}$ holds for
independent and identically distributed entries of ${\boldsymbol s}$
with $\sigma_{d}^{2}$ being the variance of the transmit signal.
Based on (\ref{eqn:sceta1}) and (\ref{eqn:prd2}), the proposed
S-SINR algorithm solves
\begin{equation}
\eta_{\varphi_{\phi_i}}^{\rm{SINR}}=\max_{i}\frac{1}{N_{i}}\sum_{n=1}^{N_{i}} \rm \frac{\sigma_{d}^{2}}{\boldsymbol h_{\phi_{in}}^{H}\boldsymbol{R}_{In}\boldsymbol h_{\phi_{in}}}.
\label{eqn:sceta1ed1}
\end{equation}
According to (\ref{eqn:sceta1ed1}), the maximization performed over $N_{i}$ data steams is difficult to achieve. To implement the optimization as expressed in (\ref{eqn:sceta1ed1}), the CSI to the relays and the correlation matrix of the interference are required. And the SINR value for each receive antenna is calculated with the obtained information. Not mention the accurate requirement of these information, the sum and invert operations for each antenna make it a sophisticated optimization algorithm. To further
simplify the maximization we assume that the streams for a device or relay have similar SINR. With this assumption we can have
\begin{equation}
\begin{split}
\eta_{\varphi_{\phi_i}}^{\rm{SINR}} & ={\min_{i}\boldsymbol h_{\phi_{\rm in}}^{H}\boldsymbol{R}_{\rm In}\boldsymbol h_{\phi_{\rm in}}}, \label{eqn:sceta1ed2} 
\end{split}
\end{equation}
We assume $\boldsymbol R_{\rm In}= \boldsymbol D_{\rm In}+\boldsymbol G$, $\boldsymbol D_{\rm In}$ is a diagonal matrix
with the diagonal elements
of $\boldsymbol R_{\rm In}$ and $\boldsymbol G$ containing the other elements
of $\boldsymbol R_{\rm In}$. With $\boldsymbol D_{\rm In}$ and $\boldsymbol G$, we can have
\begin{equation}
\begin{split}
\boldsymbol h_{\phi_{\rm in}}^{H}\boldsymbol{R}_{\rm In}\boldsymbol h_{\phi_{\rm in}} &=\boldsymbol h_{\phi_{\rm in}}^{H}\boldsymbol D_{\rm In}\boldsymbol h_{\phi_{\rm in}}
+\boldsymbol h_{\phi_{\rm in}}^{H}\boldsymbol G\boldsymbol h_{\phi_{\rm in}}\\
&=\sigma_{\rm In}^{2}\|\boldsymbol h_{\phi_{\rm in}}\|^{2}+\|\boldsymbol h_{\phi_{\rm in}}^{H}\boldsymbol G\boldsymbol h_{\phi_{\rm in}}\|,
\label{eqn:sceta1ed3}
\end{split}
\end{equation}
If we have two data streams and $\rm SINR_{1}>\rm SINR_{2}$, based on (\ref{eqn:sceta1ed1}), we can get
\begin{equation}
\boldsymbol h_{\phi_{i1}}^{H}\boldsymbol{R}_{In}\boldsymbol h_{\phi_{i1}}<\boldsymbol h_{\phi_{i2}}^{H}\boldsymbol{R}_{In}\boldsymbol h_{\phi_{i2}},
\label{eqn:sceta1ed31}
\end{equation}
with (\ref{eqn:sceta1ed3}), (\ref{eqn:sceta1ed31}) can be expressed as,
\begin{equation}
\sigma_{\rm In}^{2}\|\boldsymbol h_{\phi_{i1}}\|^{2}+\|\boldsymbol h_{\phi_{i1}}^{H}\boldsymbol G\boldsymbol h_{\phi_{i1}}\|<\sigma_{\rm In}^{2}\|\boldsymbol h_{\phi_{i2}}\|^{2}+\|\boldsymbol h_{\phi_{i2}}^{H}\boldsymbol G\boldsymbol h_{\phi_{i2}}\|,
\label{eqn:sceta1ed32}
\end{equation}
rewrite (\ref{eqn:sceta1ed32}), we can obtain
\begin{equation}
\|\boldsymbol h_{\phi_{i1}}\|^{2}<\|\boldsymbol h_{\phi_{i2}}\|^{2}+\frac{1}{\sigma_{\rm In}^{2}}(\|\boldsymbol h_{\phi_{i2}}^{H}\boldsymbol G\boldsymbol h_{\phi_{i2}}\|-\|\boldsymbol h_{\phi_{i1}}^{H}\boldsymbol G\boldsymbol h_{\phi_{i1}}\|),
\label{eqn:sceta1ed33}
\end{equation}
If $\boldsymbol G$ is small compared with ${\sigma_{\rm In}^{2}\boldsymbol I}$, we omit the term $(\|\boldsymbol h_{\phi_{i2}}^{H}\boldsymbol G\boldsymbol h_{\phi_{i2}}\|-\|\boldsymbol h_{\phi_{i1}}^{H}\boldsymbol G\boldsymbol h_{\phi_{i1}}\|)$. Finally if $\rm SINR_{1}>\rm SINR_{ 2}$,
we can have
\begin{equation}
\|\boldsymbol h_{\phi_{i1}}\|^{2}<\|\boldsymbol h_{\phi_{i2}}\|^{2},
\label{eqn:sceta1ed34}
\end{equation}
As a result, the SINR criterion can be simplified to the
selection of the channel information as described by
\begin{equation}
\hat{\eta_{\varphi_{\phi_i}}^{\rm{SINR}}}={\min_{i,n}\|\boldsymbol h_{\phi_{in}}\|^{2}}.
\label{eqn:sceta1ed4}
\end{equation}
With the criterion expressed in (\ref{eqn:sceta1ed4}), the
interference can be omitted and only the channel information is
necessary. Comparing (\ref{eqn:sceta1ed4}) and (\ref{eqn:sceta1ed1}), the optimization is performed by calculating channel gains for each antenna which is obviously easier than calculating SINRs for every antenna. Similarly, (\ref{eqn:sceta2}) can be obtained as
\begin{equation}
\hat{\eta_{\boldsymbol \varphi, r}^{\rm{SINR}}}={\min_{r,n}\|\boldsymbol h_{\phi_{in}r}\|^{2}}.
\label{eqn:sceta1ed5}
\end{equation}
Based on (\ref{eqn:sceta1ed4}) and (\ref{eqn:sceta1ed5}), the SINR
criterion in (\ref{eqn:scr}) can be simplified and the proposed
S-SINR algorithm is given by
\begin{equation}
\boldsymbol \varphi^{\rm{S-SINR}}=\rm{arg} \max_{\boldsymbol \varphi}\left(
\hat{\eta_{\rm{SINR}_{1}}},\hat{\eta_{\rm{SINR}_{2}}}\right),
\label{eqn:scrs}
\end{equation}
which only needs the channel information. In (\ref{eqn:scrs}), the calculation of $\hat{\eta_{\rm{SINR}_{1}}}$ and $\hat{\eta_{\rm{SINR}_{2}}}$ is in the same way as in (\ref{eqn:scr}).

\subsection{Simplified SR-Based (S-SR) Multiple-Relay Selection }

In the proposed S-SR algorithm with partial channel information, the
covariance matrix of the interference and the signal can be
described as $\boldsymbol R_{I}=\sum_{j\neq i}\boldsymbol
U_{j}{\boldsymbol s}_{j}^{(t)}{{\boldsymbol
s}_{j}^{(t)}}^{H}{\boldsymbol U_{j}}^{H}$ and $\boldsymbol
R_{d}=\boldsymbol U_{i}{\boldsymbol s}_{i}^{(t)}{{\boldsymbol
s}_{i}^{(t)}}^{H}{\boldsymbol U_{i}}^{H}$, respectively. If precoding matrices are assumed perfectly known in the transmission, we can further obtain an alternative way of expressing S-SR criterion. The
proposed SR-based relay selection criterion is given by
{\color{black}
\begin{align}
\boldsymbol \varphi^{\rm S-SR}&=\max_{\phi_i \in \boldsymbol \varphi, \boldsymbol \varphi \in \boldsymbol{\Psi}}
\bigg\{\log\big(\det{\left[\boldsymbol I+\boldsymbol{\Gamma}_{r,i}\right]}\big)\nonumber \\
&\qquad -\log\big(\det{\left[\boldsymbol I + \boldsymbol U_{i}^{H}\boldsymbol R_{I}^{-1}\boldsymbol U_{i}\boldsymbol R_{d}\right]}\big)\bigg\}, \label{eqn:alrnew}
\end{align}}
which can be achieved without knowledge of the channels of the
eavesdroppers. In what follows, we detail the derivation of the
S-SR relay selection algorithm.

\begin{proof}
From the original expression for the SR criterion, which is shown in
(\ref{eqn:sinrct}), we propose the following approach:
\begin{equation}
\boldsymbol{\varphi}=\rm{arg} \max_{\phi_i \in \boldsymbol \varphi,\boldsymbol \varphi \in \boldsymbol{\Psi}} \left\{
\frac{\det(\boldsymbol I +{\boldsymbol{\Gamma}_{r,i}})}{\det(\boldsymbol I+{\boldsymbol{\Gamma}_{e,i}})}\right\},
\label{eqn:alsinrct4}
\end{equation}
{\color{black}In (\ref{eqn:alsinrct4}), our aim is to eliminate the channel information of eavesdroppers from the denominator. With square matrices, $\det(\boldsymbol
AB)=\det(\boldsymbol A)\det(\boldsymbol B)$, to satisfy the requirement, the denominator can be expressed as
\begin{equation}
{\det[{\boldsymbol \Lambda_1}^{-1}{\boldsymbol \Lambda_1}+{\boldsymbol \Lambda_1}^{-1}(\boldsymbol H_{e}\boldsymbol R_{d}\boldsymbol
H_{e}^{H})]}, \label{eqn:alsinrct411}
\end{equation}
where ${\boldsymbol \Lambda_1}=\boldsymbol H_{e}\boldsymbol R_{I}\boldsymbol H_{e}^{H}$. As $\boldsymbol \Lambda_1$ is a square matrix, (\ref{eqn:alsinrct411}) can be obtained as,
\begin{equation}
\det[{\boldsymbol \Lambda_1}^{-1}]{\det[{\boldsymbol \Lambda_1}+(\boldsymbol H_{e}\boldsymbol R_{d}\boldsymbol
H_{e}^{H})]}, \label{eqn:alsinrct412}
\end{equation}}
Using the property of the determinant $\det(\boldsymbol
A^{-1})=\frac{1}{\det(\boldsymbol A)}$ \cite{matrix}, we have
\begin{equation}
\det[{\boldsymbol \Lambda_1}]^{-1}{\det[{\boldsymbol \Lambda_1}+(\boldsymbol H_{e}\boldsymbol R_{d}\boldsymbol
H_{e}^{H})]},
\label{eqn:alsinrct42}
\end{equation}
In (\ref{eqn:alsinrct42}), we separate the equation into two parts.
\begin{align}
\det[{\boldsymbol \Lambda_1}]^{-1}&=\det[\boldsymbol H_{e}\boldsymbol R_{I}\boldsymbol H_{e}^{H}]^{-1} \nonumber \\
&=\det[\boldsymbol H_{e}(\sum_{j\neq i}\boldsymbol
U_{j}{\boldsymbol s}_{j}^{(t)}{{\boldsymbol
s}_{j}^{(t)}}^{H}{\boldsymbol U_{j}}^{H})\boldsymbol H_{e}^{H}]^{-1}
\label{eqn:alsinrct421}
\end{align}
on the left-hand side of equation (\ref{eqn:alsinrct421}), we multiply $\boldsymbol
U_{i}{\boldsymbol U_{i}}^{-1}=\boldsymbol I$ and on the right side we multiply ${\boldsymbol U_{i}^{H}}^{-1}{\boldsymbol U_{i}}^{H}=\boldsymbol I$, we can have,
\begin{equation}\begin{split}
\det[{\boldsymbol \Lambda_1}]^{-1}&=\{\det[\boldsymbol H_{e}\boldsymbol
U_{i}] \\
&\det[(\sum_{j\neq i}{\boldsymbol
U_{i}}^{-1}\boldsymbol
U_{j}{\boldsymbol s}_{j}^{(t)}{{\boldsymbol
s}_{j}^{(t)}}^{H}{\boldsymbol U_{j}}^{H}{\boldsymbol
U_{i}^{H}}^{-1})] \\
&\det[{\boldsymbol
U_{i}^{H}}\boldsymbol H_{e}^{H}]\}^{-1}, \label{eqn:alsinrct423}
\end{split}\end{equation}
Similarly, the second part in equation (\ref{eqn:alsinrct42}) can be obtained as,
\begin{equation}\begin{split}
&{\det[{\boldsymbol \Lambda_1}+(\boldsymbol H_{e}\boldsymbol R_{d}\boldsymbol
H_{e}^{H})]}=\det[\boldsymbol H_{e}\boldsymbol
U_{i}] \\
&\det[(\sum_{j\neq i}{\boldsymbol
U_{i}}^{-1}\boldsymbol
U_{j}{\boldsymbol s}_{j}^{(t)}{{\boldsymbol
s}_{j}^{(t)}}^{H}{\boldsymbol U_{j}}^{H}{\boldsymbol U_{i}^{H}}^{-1})+{\boldsymbol s}_{i}^{(t)}{{\boldsymbol
s}_{i}^{(t)}}^{H}]\\
&\det[{\boldsymbol
U_{i}^{H}}\boldsymbol H_{e}^{H}], \label{eqn:alsinrct425}
\end{split}\end{equation}

As the matrices $\boldsymbol H_{e}\boldsymbol
U_{i}$ and ${\boldsymbol
U_{i}^{H}}\boldsymbol H_{e}^{H}$ are square and have equal size, based on equation (\ref{eqn:alsinrct42}), (\ref{eqn:alsinrct423}) and (\ref{eqn:alsinrct425}) we can
eliminate the term $\det[\boldsymbol H_{e}\boldsymbol
U_{i}]$ and $\det[{\boldsymbol
U_{i}^{H}}\boldsymbol H_{e}^{H}]$. The secrecy rate selection criterion (\ref{eqn:alsinrct4}) can be rewrite as,
\begin{equation}
\boldsymbol{\varphi}=\rm{arg} \max_{\phi_i \in \boldsymbol \varphi,\boldsymbol \varphi \in \boldsymbol{\Psi}}
\bigg\{\frac{\det[\boldsymbol I+\boldsymbol{\Gamma}_{r,i}]}{\det[{\boldsymbol \Lambda_2}]}\bigg\}
\label{eqn:alsinrct431}
\end{equation}
where ${\boldsymbol \Lambda_2}$ require only the information of precoding matrices and transmit symbols which is expressed as:
\begin{equation}\begin{split}
&{\boldsymbol \Lambda_2}=(\sum_{j\neq i}{\boldsymbol
U_{i}}^{-1}\boldsymbol
U_{j}{\boldsymbol s}_{j}^{(t)}{{\boldsymbol
s}_{j}^{(t)}}^{H}{\boldsymbol U_{j}}^{H}{\boldsymbol
U_{i}^{H}}^{-1})^{-1}  \\
&[(\sum_{j\neq i}{\boldsymbol
U_{i}}^{-1}\boldsymbol
U_{j}{\boldsymbol s}_{j}^{(t)}{{\boldsymbol
s}_{j}^{(t)}}^{H}{\boldsymbol U_{j}}^{H}{\boldsymbol U_{i}^{H}}^{-1})+{\boldsymbol s}_{i}^{(t)}{{\boldsymbol
s}_{i}^{(t)}}^{H}]\label{eqn:alsinrct4311}
\end{split}\end{equation}
with $\boldsymbol R_{I}$ and $\boldsymbol R_{d}$, we can have
\begin{equation}
{\boldsymbol \Lambda_2}=\boldsymbol I + \boldsymbol U_{i}^{H}\boldsymbol R_{I}^{-1}\boldsymbol U_{i}\boldsymbol R_{d},
\label{eqn:alsinrct441}
\end{equation}
By adding $\log$ to (\ref{eqn:alsinrct441}), we obtain
\begin{equation}
\boldsymbol \varphi^{\rm S-SR}=\rm{arg} \max_{\phi_i \in \boldsymbol \varphi,\boldsymbol \varphi \in \boldsymbol{\Psi}} \left\{\log\Big(
\frac{\det[\boldsymbol I+\boldsymbol{\Gamma}_{r,i}]}{\det[\boldsymbol I + \boldsymbol U_{i}^{H}\boldsymbol R_{I}^{-1}\boldsymbol U_{i}\boldsymbol R_{d}]}\Big)\right\},
\label{eqn:alsinrct451}
\end{equation}

which is equivalent to (\ref{eqn:alrnew}). In the derivation,
we assume the channel matrices of the users
have the same matrix size as the channel of the eavesdroppers and
the matrices are full rank.
\end{proof}


\section{Simulation Results}

In this section, we assess the secrecy rate performance in a
multiuser MIMO downlink relay system. In the
simulation, 5 relays are placed between the source and the users. Zero-forcing precoding is adopted and we assume that the
channel for each user is uncorrelated with the remaining channels
and the channel gains are generated following a complex circular
Gaussian random variable with zero mean and unit variance.

%

\begin{figure}[ht]
\centering
\includegraphics[width=\linewidth]{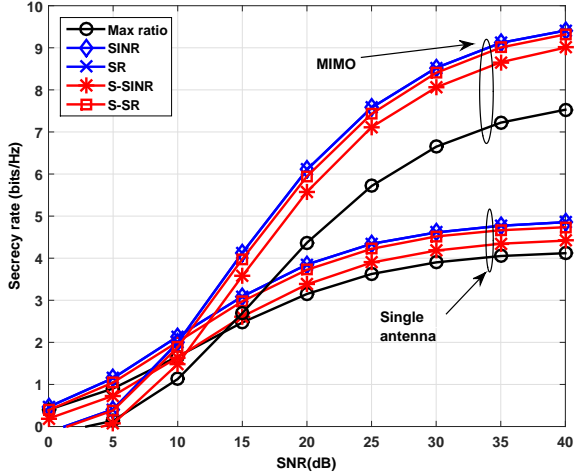}
\vspace{-0.5em} \caption{Single antenna and MIMO multi-user system
secrecy rate performance with different relay selection criteria}
\label{fig:pa2}
\end{figure}

In Fig. \ref{fig:pa2}
single-antenna scenario, the SINR and SR criteria have a comparable
SR performance. However, the SINR criterion requires the
interference knowledge and the SR criterion needs the eavesdroppers
channel which are both impractical in downlink transmissions. The
proposed S-SR algorithm only requires the channels to the relays and
the legitimate users and can achieve almost the same SR performance
as the SR criterion with full channel knowledge. The proposed S-SINR
algorithm suffers a larger degradation than that of the SR
criterion.

\begin{figure}[ht]
\centering
\includegraphics[width=\linewidth]{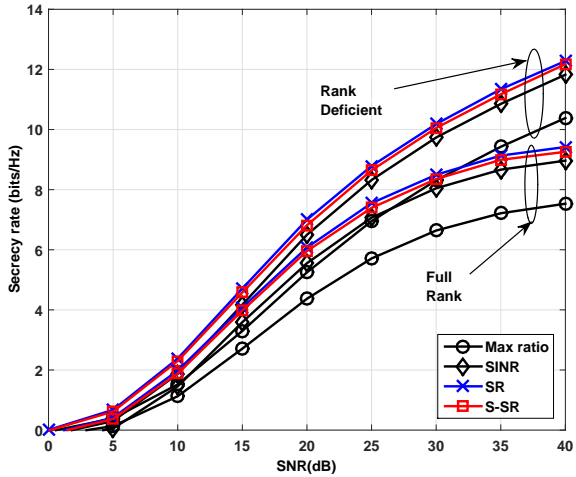}
\caption{Secrecy rate performance with different relay selection
criteria in full-rank and rank-deficient systems} \label{fig:pa3}
\end{figure}

In Fig. \ref{fig:pa3}, we compare the SR performance for scenarios
with square and equal channel matrices and that with the zeros
filled out. When we decrease the number of eavesdroppers, the SR
performance increases. In this scenario, the proposed S-SR relay
selection algorithm can still perform close to the SR relay
selection criterion with full information.

\begin{figure}[ht]
\centering
\includegraphics[width=\linewidth]{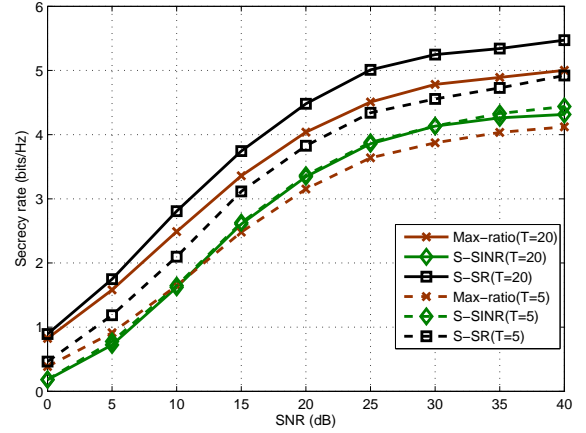}
\caption{Secrecy rate performance and the effect of the numbers of
relays using single-antenna devices with simplified CSI knowledge
criteria.} \label{fig:pa5}
\end{figure}

In Figure~\ref{fig:pa5}, simplified criteria are compared along with
different numbers of relays. The S-SR criterion provides the best
secrecy rate performance among all investigated criteria. With more
relays distributed between the source and users, the secrecy rate
performance can be further improved by employing the proposed S-SR
and S-SINR criteria.

\begin{figure}[ht]
\centering
\includegraphics[width=\linewidth]{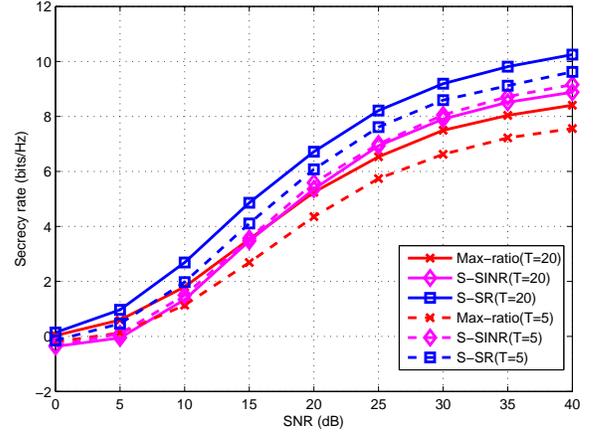}
\caption{Secrecy rate performance and the effect of the numbers of
relays using multiple-antenna devices with simplified CSI knowledge
criteria.} \label{fig:pa5}
\end{figure}

\section{Conclusion}

In this work, we have proposed effective multiple-relay selection
algorithms for multiuser MIMO relay systems to enhance the
legitimate users' transmission. The proposed algorithms exploit the
use of the available channel information to perform relay selection.
Simulation results show that the proposed algorithms can provide a
significantly better secrecy rate performance than existing
approaches.


%



\ifCLASSOPTIONcaptionsoff
  \newpage
\fi

\end{document}